\documentclass[%
 reprint,
superscriptaddress,
showpacs,preprintnumbers,
notitlepage,
 amsmath,amssymb,
 aps,
prx,
floatfix,
]{revtex4-2}

\usepackage{graphicx}
\usepackage{xcolor}
\usepackage{bm}
\usepackage{subfig}
\newsavebox{\measurebox}

\begin{document}


\title{Simplicial contagion in temporal higher-order networks}

\author{Sandeep Chowdhary}
\affiliation{Department of Network and Data Science, Central European University, Vienna}
\author{Aanjaneya Kumar}
\affiliation{Department of Physics, Indian Institute of Science Education and Research, Dr. Homi Bhabha Road, Pune 411008, India}
\author{Giulia Cencetti}
\affiliation{Fondazione Bruno Kessler, Trento, Italy}
\author{Iacopo Iacopini}
\affiliation{Aix Marseille Univ, Universit\'e de Toulon, CNRS, CPT, Marseille, France}
\author{Federico Battiston}
\affiliation{Department of Network and Data Science, Central European University, Vienna}


\begin{abstract}
Complex networks represent the natural backbone to study epidemic processes in populations of interacting individuals. Such a modeling framework, however, is naturally limited to pairwise interactions, making it less suitable to properly describe social contagion, where individuals acquire new norms or ideas after simultaneous exposure to multiple sources of infections. Simplicial contagion has been proposed as an alternative framework where simplices are used to encode group interactions of any order. The presence of higher-order interactions leads to explosive epidemic transitions and bistability which cannot be obtained when only dyadic ties are considered. In particular, critical mass effects can emerge even for infectivity values below the standard pairwise epidemic threshold, where the size of the initial seed of infectious nodes determines whether the system would eventually fall in the endemic or the healthy state. Here we extend simplicial contagion to time-varying networks, where pairwise and higher-order simplices can be created or destroyed over time. By following a microscopic Markov chain approach, we find that the same seed of infectious nodes might or might not lead to an endemic stationary state, depending on the temporal properties of the underlying network structure, and show that persistent temporal interactions anticipate the onset of the endemic state in finite-size systems. We characterize this behavior on higher-order networks with a prescribed temporal correlation between consecutive interactions and also on heterogeneous simplicial complexes, showing that temporality again limits the effect of higher-order spreading, but in a less pronounced way than for homogeneous structures. Our work suggests the importance of incorporating temporality, a realistic feature of many real-world systems, into the investigation of dynamical processes beyond pairwise interactions.
\end{abstract}

\maketitle

\section{Introduction}
Contagion processes, from the spread of diseases to opinions and rumors, are ubiquitous in nature~\citep{keeling2011modeling,goffman1964generalization, daley1964epidemics}. In all such cases, the contact structure of the underlying population  has a crucial role in determining the emerging collective behavior, making network science one of the primary tools to investigate spreading dynamics in real-world systems~\citep{boccaletti2006complex, barrat2008dynamical, pastorsatorras2015epidemic, porter2016dynamical, arruda2018fundamentals}. For instance, pioneering investigations have shown that heavy-tailed degree distributions in the contact structure lead to a vanishing epidemic threshold, a behavior which can not be observed neither in well-mixed population nor in homogeneous networks~\cite{pastorsatorras2001epidemic}. For the biological spread of pathogens, contagion is typically mediated by pairwise interactions, where each link represents an independent source of infection. However, this mechanism of {\it simple} contagion does not seem to accurately describe social contagion. To acquire new ideas, norms or opinions, spreading is better modelled by {\it complex} contagion~\cite{centola2007complex,watts2007influentials,lehmann2018complex,watts2002simple}. In this case, individuals are subject to the simultaneous pressure of their neighbors, leading to a dynamics of cascades which has also been empirically observed in a number of different contexts~\cite{centola2010spread,ugander2012structural, weng2012competition,karsai2014complex,monsted2017evidence}. 

For many years, the wide majority of networked systems have been represented by graphs, collection of edges and links, where interactions are naturally limited to dyadic ones. However, in most real-world networks, interactions can also occur among groups composed by three or more individuals. All these systems are better described by simplicial complexes or hypergraphs, which naturally take into account the presence of higher-order interactions, providing a suitable extension of the traditional network framework beyond pairwise interactions~\cite{lambiotte2019networks, battiston2020networks, torres2020why, bick2021what}. 
In particular, {\it simplicial} contagion is a newly proposed paradigm that allows to model at the microscopic scale the effect of group interactions (described as simplices of different order) on spreading dynamics~\cite{iacopini2019simplicial}. Interestingly, if the infection rate associated to the higher-order interactions is high enough, this leads to the emergence of new collective behavior, making the transition from the healthy to the endemic phase explosive, and giving rise to metastable states. This result, obtained analytically by a mean-field analysis and confirmed by numerical simulations~\cite{iacopini2019simplicial, barrat2021social}, has also been replicated under different modeling frameworks, such as the microscopic Markov chain approach and the generalised link equation~\cite{matmalas2020abrupt, burgio2021network}, and on different higher-order representations, such as hypergraphs~\cite{arruda2020social, landry2020effect, arruda2021phase}. The disruptive presence of higher-order interactions is not limited to contagion dynamics, as new collective behavior has also been observed in the case of synchronization phenomena~\cite{bick2016chaos,skardal2019abrupt, millan2020explosive, lucas2020multiorder}, random walk~\cite{carletti2020random,schaub2020random}, consensus~\cite{neuhauser2020multibody, iacopini2021vanishing}, ecological~\cite{bairey2016high, grilli2017higher} and evolutionary dynamics~\cite{alvarez2021evolutionary} when extended beyond simple dyadic ties. 
For pairwise contagion, the temporal nature of interactions, where links can be created and destroyed over time, is known to significantly affect the evolution and the long-term properties of the spreading process~\cite{rocha2011simulated, karsai2011small}. Indeed, temporal networks~\cite{holme2012temporal} are routinely used as a modeling framework to properly capture diffusive processes taking place on realistic populations where the contact structure changes over time ~\cite{perra2012activity,valdano2015analytical,masuda2017temporal,koher2019contact}. Recently, also higher-order social networks have been found to have a non-trivial temporal dynamics~\cite{cencetti2021temporal}. Yet, so far very little attention has been devoted to understanding how temporality affects spreading on higher-order structures~\citep{st-onge2021bursty}. 

Here, we extend models of simplicial contagion to the case of time-varying networks, where both pairwise and higher-order interactions can evolve over time. We compare the contagion process on static and temporal simplicial complexes.
The dynamics of the static case presents bistability, meaning that the long-term behavior of the system is determined by the size of the initial seed of infectious nodes. In our work, we numerically characterize the basins of attraction of healthy and endemic states in static and temporal higher-order structures, showing that persistent temporal interactions anticipate the onset of the endemic state in finite-size systems. This means that the same number of initially infected agents might or might not lead to an endemic stationary state, depending on the temporal properties of the underlying network structure. To this aim, we propose a simple model to tune the degree of temporal correlations in synthetic structures that evolve over time, and investigate how this variable affects the long-term outcome of the spreading dynamics. We show that temporality can significantly reduce the enhancement of epidemics typically induced by higher-order contagion terms in the forward transition to the endemic state. By contrast, the backward transition to the infection-free state remains unaffected by presence of temporal correlation or lack thereof. Finally, we study simplicial contagion on temporal higher-order networks that present degree heterogeneity, showing once again that temporality hinders higher-order spreading, but in a less pronounced way than for homogeneous structures.

 \begin{figure*}[bt]
    \centering
\begin{minipage}{1\textwidth}
\includegraphics[scale=0.3]{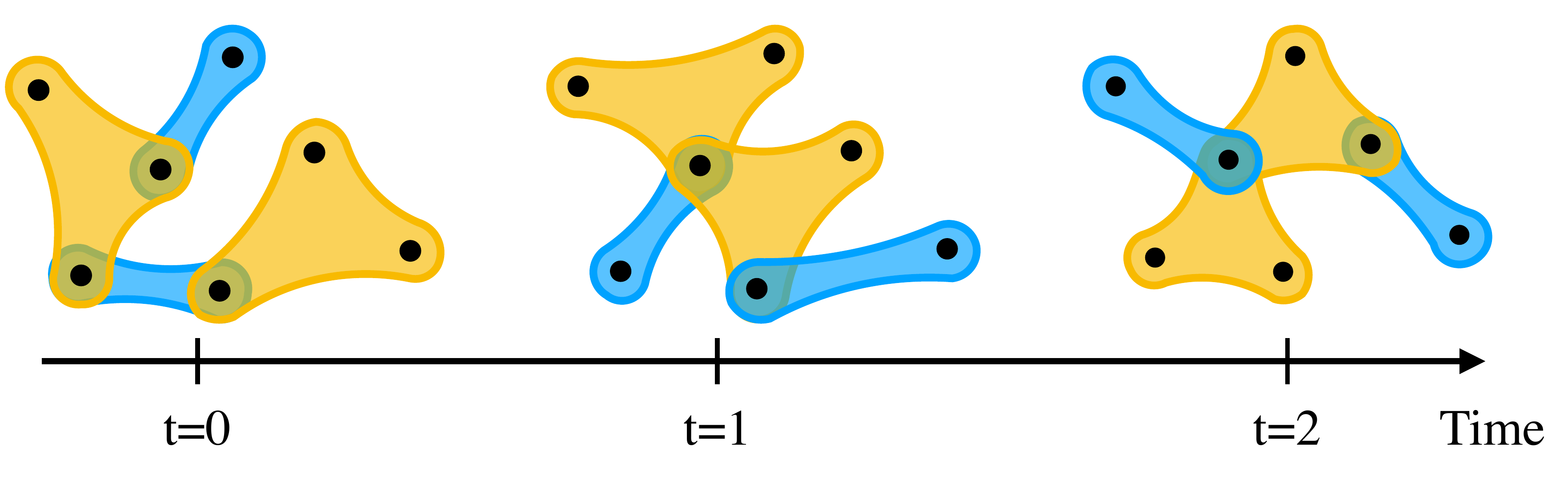}
\end{minipage}
     \caption{\textbf{Temporal higher-order networks.} Schematic of a time-varying higher-order network where both pairwise and higher-order interactions evolve over time.}
    \label{fig:fig1_schematic}
\end{figure*} 

\section{Model}

We study social contagion in simplicial complexes which evolve over time. In particular, following Ref.\cite{iacopini2019simplicial}, we consider an SIS model, where each one of the $N$ interacting nodes can be in either of two states -- susceptible ($S$) or infected ($I$). We consider interactions up to groups of three, such that 1-simplices (links) encode standard pairwise interactions, while 2-simplices describe three individuals interacting together (and this is structurally different from having three links that form a triangle). 

In a time-step of the SIS model, any infected individual can infect their susceptible neighbours connected by 1-simplices with a probability $\beta_|$, and infected nodes can recover with probability $\mu$ and become susceptible again. However, in the simplicial version of the model, 2-simplices provide an additional way for a contagion event to happen. In particular, if a susceptible individual is part of a 2-simplex while the other two members of the simplex are infected, there is an additional probability $\beta_\triangle$ to also get infected -- associated to a microscopic description of social reinforcement induced by group interactions.
 
 We write the discrete time evolution equation for the infection probabilities of each node at a particular instant using the Microscopic Markov Chain Approach (MMCA)~\citep{gomez2010discrete}. MMCAs have been extended to temporal networks, allowing for an analytical computation of the epidemic threshold~\cite{valdano2015analytical}, and more recently to simplicial complexes, though in this context the non-linear term associated to contagion in 2-simplices only allows a numerical solution~\cite{matmalas2020abrupt}. According to this approach, the probability of a generic node $i$ to be infected at time $t+1$ is 
\begin{multline}\label{eq:pi}
      p_i(t + 1) = (1-q_i(t)q_{i,\triangle}(t))(1-p_i(t)) + (1-\mu)p_i(t),
\end{multline}
where the first term on the right-hand side of Eq.~\eqref{eq:pi} represents the probability at time $t$ for a susceptible node to get infected. This is given by the product of $(1-p_i(t))$, the probability that node $i$ is susceptible, and $(1-q_i(t)q_{i,\triangle}(t))$, the probability that $i$ is infected by at least one of its neighbours. The second term, $(1-\mu)p_i(t)$, stands for the probability that node $i$ is already infected at time $t$ and does not recover. 
Here $q_i(t)$ defines the probability that node~$i$ is not infected via pairwise interactions with its neighbours,
\begin{equation}
  \label{eq:q_i_mmca}
  q_i(t) = \prod_{j\in\Gamma_i(t)}\left( 1 - \beta_| p_j(t) \right)\,,
\end{equation}
with $\Gamma_i(t)$ denoting the set of 1-simplices containing node $i$ at time $t$. Similarly, $q_{i,\triangle}(t)$ defines the probability that node~$i$ is not infected by any of its 2-simplicial interactions, 
\begin{equation}
  \label{eq:q_i_tri_mmca}
  q_{i,\triangle}(t) = \prod_{j, \ell \in \triangle_i(t)}\left( 1 - \beta_\triangle p_j(t)p_\ell(t) \right)\,,
\end{equation}
with $\triangle_i(t)$ denoting the set of 2-simplices containing node $i$ at time $t$.

Notice how, in contrast with Ref.~\cite{matmalas2020abrupt}, here $\Gamma_i(t)$ and $\Delta_i(t)$ are functions of time, and allow us to generalize the MMCA approach to evolving simplicial complexes.

\section{Results}

\subsection{Social contagion on static and temporal simplicial complexes}

\begin{figure*}[!htb]
    \centering
    \begin{minipage}{.4\textwidth}
        \subfloat []
        {\label{fig:heat_a}\includegraphics[scale=0.45]{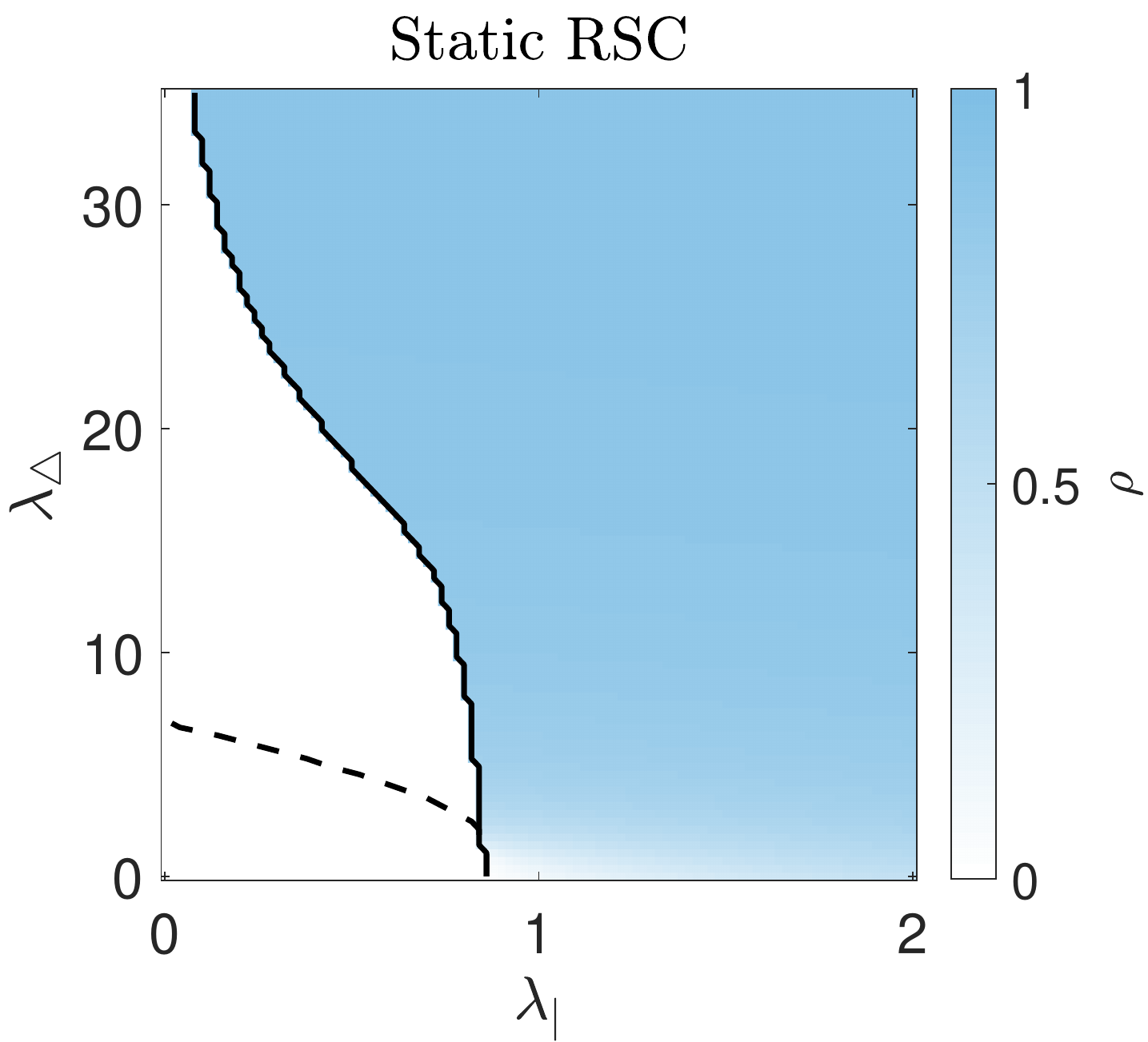}}
    \end{minipage}%
    \begin{minipage}{0.4\textwidth}
        \subfloat []
        {\label{fig:heat_b}\includegraphics[scale=0.45]{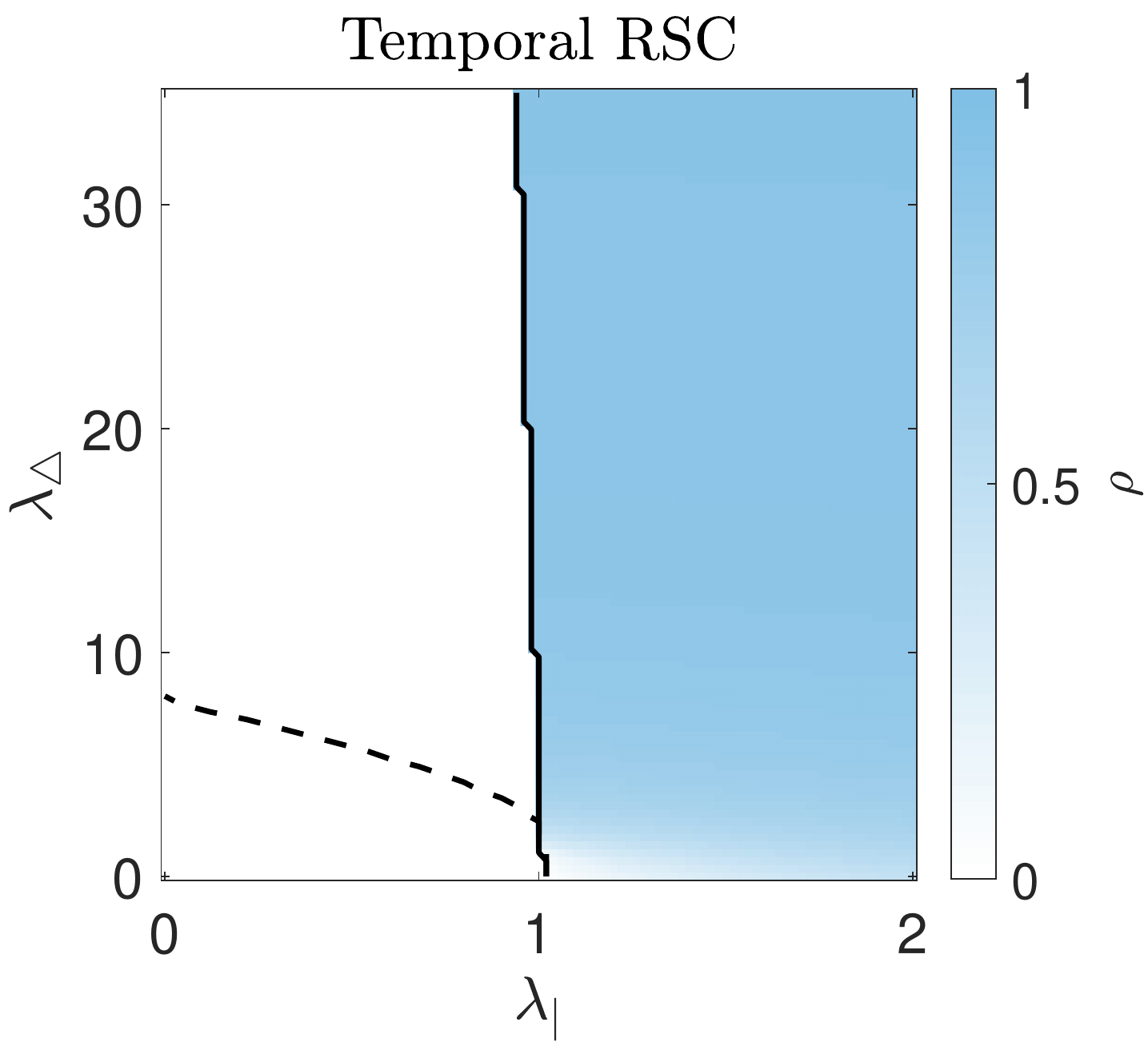}}
    \end{minipage}
    \caption{\textbf{Contagion on static and temporal simplicial complexes.} We show the fraction of infected nodes at the equilibrium starting from a single infected node as a function of rescaled pairwise $\lambda_|$ and simplicial $\lambda_\triangle$ infection rates for static (a) and temporal (b) simplicial complexes with $N=500$ nodes. In the static case, the epidemic onset (solid black line) as a function of $\lambda_|$ is anticipated as we increase $\lambda_{\triangle}$. This suggests that the chosen initial infection of size $\frac{1}{N}$ belongs to the basin of the infection-free state for small values of $\lambda_{\triangle}$, moving into the basin of the endemic state upon increasing $\lambda_{\triangle}$. For time-evolving higher-order networks such effect is not observed, and we find a suppression of the endemic phase which can not be reached for low values of $\lambda_|$, independently on the value of $\lambda_\triangle$. The backward transition to the infection-free state (dashed black lines) is largely unaffected by the temporality of the interactions. We set  $\mu=0.1$, $\langle{k_|}\rangle=12$ and $\langle{k_\triangle}\rangle=5$ for both scenarios.}
    \label{fig:heat}
\end{figure*}

We begin by comparing contagion processes in static simplicial complexes and in higher-order networks that change over time. A schematic of a time-varying higher-order network is shown in Fig.\ref{fig:fig1_schematic} where 1-simplices and 2-simplices are respectively coloured in blue and yellow. 
In particular, we consider random simplicial complexes (RSCs) with $N=500$ nodes generated following the algorithm introduced in Ref.~\citep{iacopini2019simplicial}. The procedure allows to obtain homogeneous simplicial complexes with controlled generalised degree properties~\cite{courtney2016generalized}, namely $\langle{k_|}\rangle$, the standard pairwise degree, and $\langle k_\triangle\rangle$, the average number of 2-simplices incident on a node.
In such model, 1-simplices are created akin to the Erdös-Rényi model, by connecting any pair $(i, j)$ of vertices with
probability $p_|$. Similarly, 2-simplices are added by connecting any triplet $(i, j, \ell)$ of vertices with probability $p_\triangle$. For two desired
values of $\langle{k_|}\rangle$ and $\langle k_\triangle\rangle$ it is possible to choose $p_|$ and $p_\triangle$ according to:
$p=\frac{\langle{k_|}\rangle - 2 \langle k_\triangle\rangle}{N-1-2\langle k_\triangle\rangle}$
and $p_\triangle=\frac{2\langle k_\triangle\rangle}{(N-1)(N-2)}$ \citep{iacopini2019simplicial}. 

We are particularly interested in studying how temporality affects the basins of attraction in the bistable regime which separate the endemic state from the infection-free state. Thus, we simulate the contagion process by first infecting a single node chosen at random and check whether this is sufficient or not to fall into the absorbing state with no epidemics. In particular, we numerically track the temporal evolution of the system at each time step $t$ by updating the infection probabilities $p_i(t)$ for all nodes as dictated by Eq.~\eqref{eq:pi}. We iterate Eq.~\eqref{eq:pi} for long time (10000 time steps) and compute the density of infected node in the stationary state by averaging the infection probabilities as $\rho=\frac{\sum_i p_i}{N}$.

In Fig.~\ref{fig:heat_a} we show $\rho$ for a static RSC as a function of rescaled pairwise, $\lambda_| = \beta_| \frac{\langle{k_|}\rangle}{\mu}$, and simplicial, $\lambda_\triangle  = \beta_\triangle \frac{\langle{k}_\triangle\rangle}{\mu}$ infection parameters. In Fig.~\ref{fig:heat_b} we compute $\rho$ for RSCs that change over time, where at each time $t$ we generate a new realisation of the RSC model with the same $\langle{k_|}\rangle$ and $\langle{k_\triangle}\rangle$ of the static simulations. In both heatmaps, two distinct regions separated by the black solid curves appear, an infection-free region where $\rho=0$ and an endemic region where a macroscopic fraction of the nodes is infected.

\begin{figure*}[!htb]
    \centering
    \begin{minipage}{.32\textwidth}
        \subfloat []
        {\label{fig:finite_a}\includegraphics[scale=0.45]{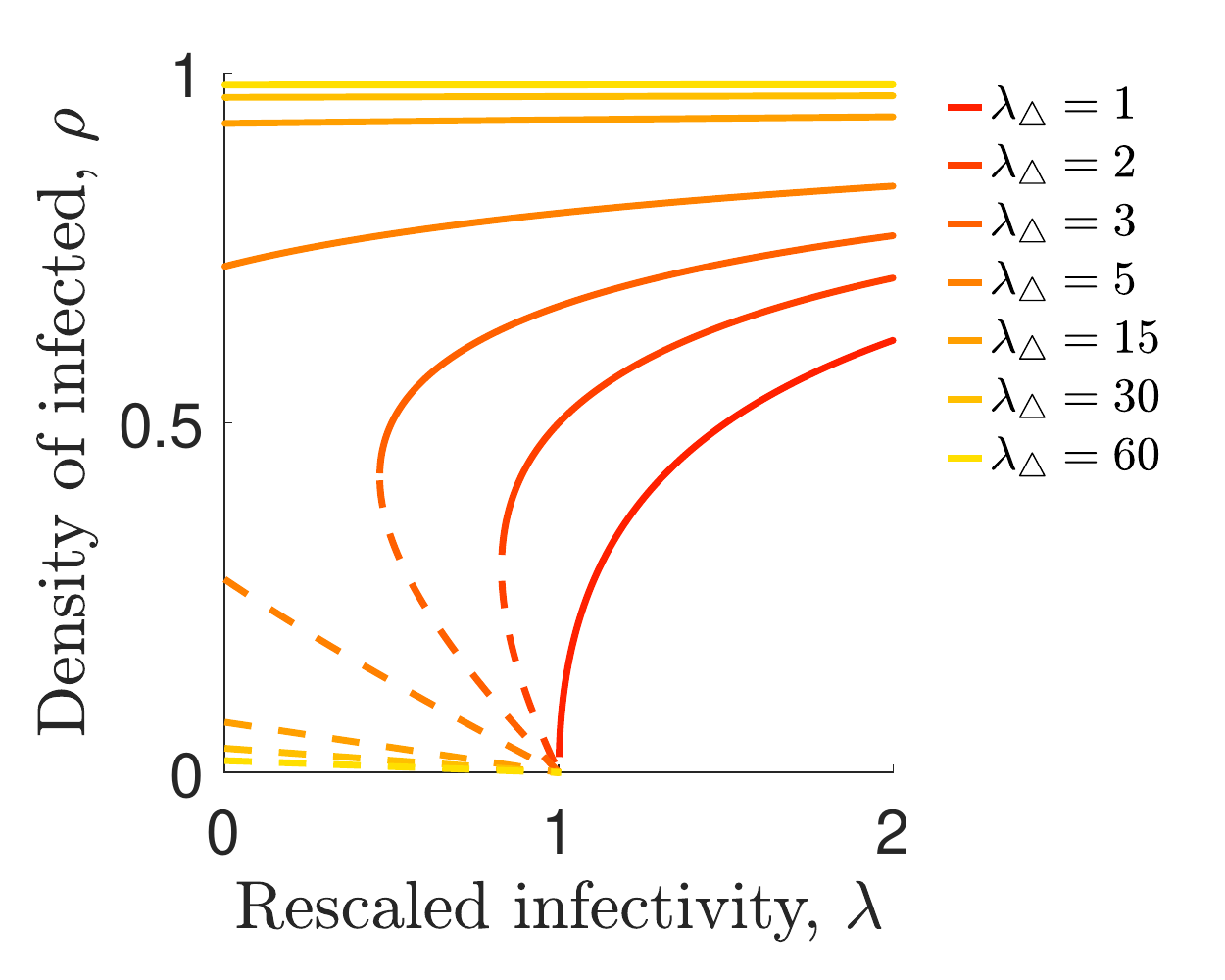}}
    \end{minipage}%
    \begin{minipage}{0.32\textwidth}
        \subfloat []
        {\label{fig:finite_b}\includegraphics[scale=0.45]{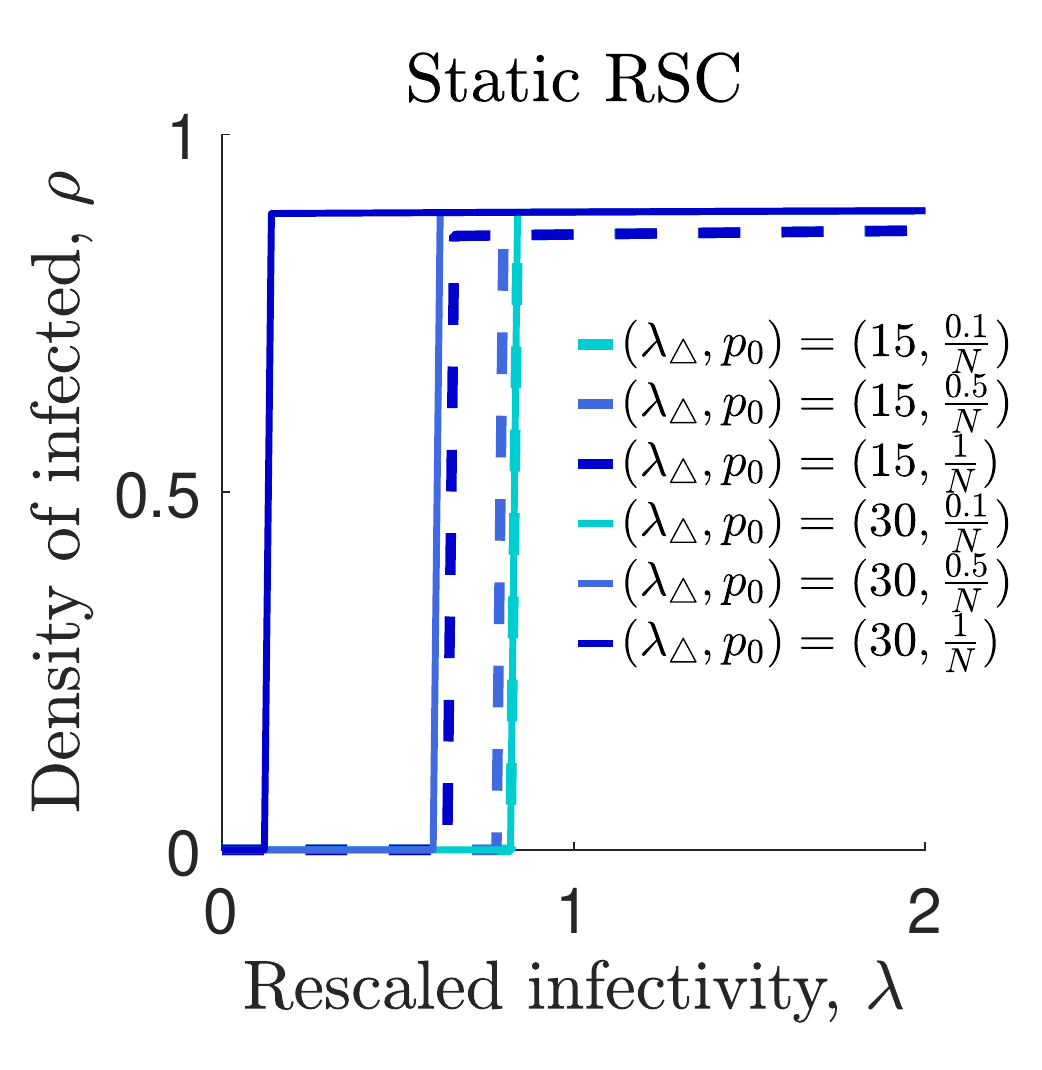}}
    \end{minipage}
    \begin{minipage}{.32\textwidth}
        \subfloat []
        {\label{fig:finite_c}\includegraphics[scale=0.45]{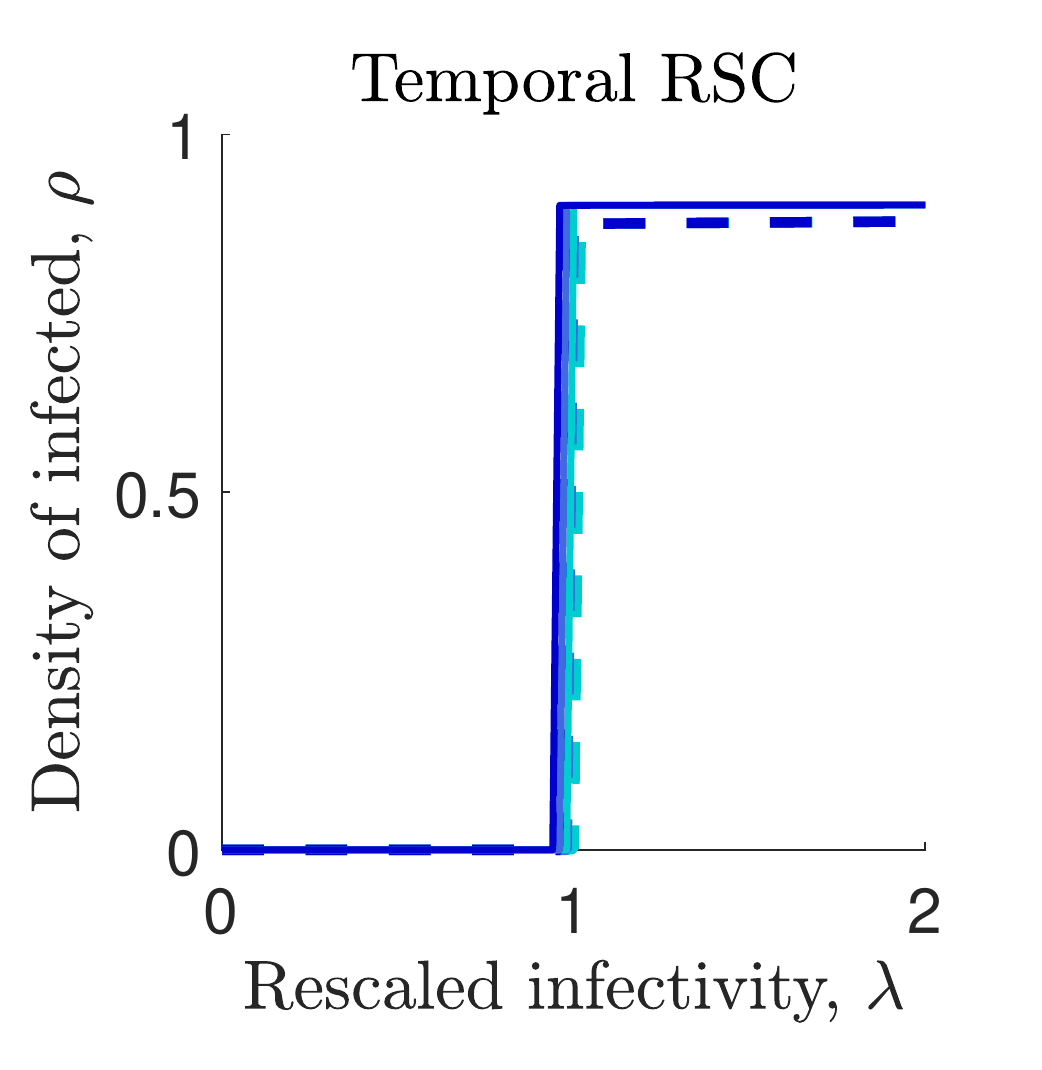}}
    \end{minipage}%
    \caption{\textbf{Effect of initial infection size on the onset of the endemic state.} (a): Size of the infected population as function of $\lambda_|$  obtained analytically in the mean-field limit for different values of $\lambda_\triangle$. The basin of the infection-free state shrinks as $\lambda_\triangle$ is increased allowing earlier onset of endemic phase for the forward transition. 
    Density of infected nodes for static (b) and temporal (c) simplicial complexes as function of $\lambda_|$ for three different initial infections, $p_0=\frac{0.1}{N}, \frac{0.5}{N}, \frac{1}{N}$ and two different values of rescaled simplicial infectivity, $\lambda_\triangle=15$ (dashed curves) and $\lambda_\triangle=30$ (solid curves). An early onset of the endemic phase is observed for sufficiently high values of the infected seeds and $\lambda_\triangle$ with a MMCA approach, compatible with our observations in (a). By contrast, in temporal simplicial complexes, even for higher values of initial infection ~ $\frac{1}{N}$ and high simplicial infectivity, (e.g.$\lambda_\triangle=30$), there is a striking suppression of contagion and early onset of endemic state does not occur. We set  $\mu=0.1$, $\langle{k_|}\rangle=12$ and $\langle{k_\triangle}\rangle=5$ for both static and temporal scenarios. }
    \label{fig:finite}
\end{figure*}

In the static case, as we increase $\lambda_{\triangle}$, the epidemic onset occurs for progressively smaller values of $\lambda_|$ in finite-size systems. This means that the seed of infectious nodes of fixed size $\frac{1}{N}$ belongs to the basin of attraction of the infection-free state for small values of $\lambda_{\triangle}$, while it moves to the basin of the endemic state upon increasing $\lambda_{\triangle}$. Coherently with the results obtained with the mean-field formalism~\cite{iacopini2019simplicial}, above a critical value of $\lambda_|$ ($\lambda_|=1$ in the mean-field approximation) the system always reaches a non-zero fraction of infected agents which grows together with $\lambda_{\triangle}$. More interestingly, before this critical value, it is still possible to end up in the endemic state due to the higher-order contributions, but only if the seed of infectious nodes is big enough (critical mass). In this case, the system undergoes an abrupt transition.

Surprisingly, by contrast, $\lambda_\triangle$ does not affect the onset of the epidemics in temporal simplicial complexes of finite size. This is clear from Fig.~\ref{fig:heat_b}, where the transition from the healthy to the endemic state is only observed as a function of $\lambda_|$, with the critical point $\lambda_|^c=1$ coinciding with what predicted by the mean-field approach~\cite{iacopini2019simplicial}. Notice indeed that critical mass effects are completely suppressed, and below $\lambda_|^c$ the same seed of infectious nodes can never sustain the epidemics --as opposed to what happens in the static case for sufficiently high values of $\lambda_\Delta$.

It is worth mentioning that in static structures [Fig.~\ref{fig:heat_a}] we also find a slight anticipation of the epidemic threshold due to the MMCA as compared to the mean-field treatment, according to which the critical threshold $\lambda^c_|=1$ for $\lambda_\triangle=0$. This is consistent with what has been already observed in Refs.~\cite{matmalas2020abrupt, burgio2021network}.\\

So far we have focused on forward transitions from the infection-free state to the endemic state. Yet, abrupt transitions are typically associated to the emergence of hysteresis cycles. For this reason we also explore the backwards transition from the endemic phase to the infection-free state by choosing the stationary-state infection probabilities obtained at the higher value of $\lambda_|$ as the initial seeds for simulations at lower $\lambda_|$ values. We show the backward transitions as dashed black lines in Fig.~\ref{fig:heat_a}, \ref{fig:heat_b} and find that they remain unaffected by temporality.\\

\begin{figure*}
\centering
\sbox{\measurebox}{%
  \begin{minipage}[b]{.4\textwidth}
  \vspace{0.6cm}
  \subfloat
    []
    {\label{fig:corr_a}\hspace*{-0.6cm}\includegraphics[width=1.1\textwidth]{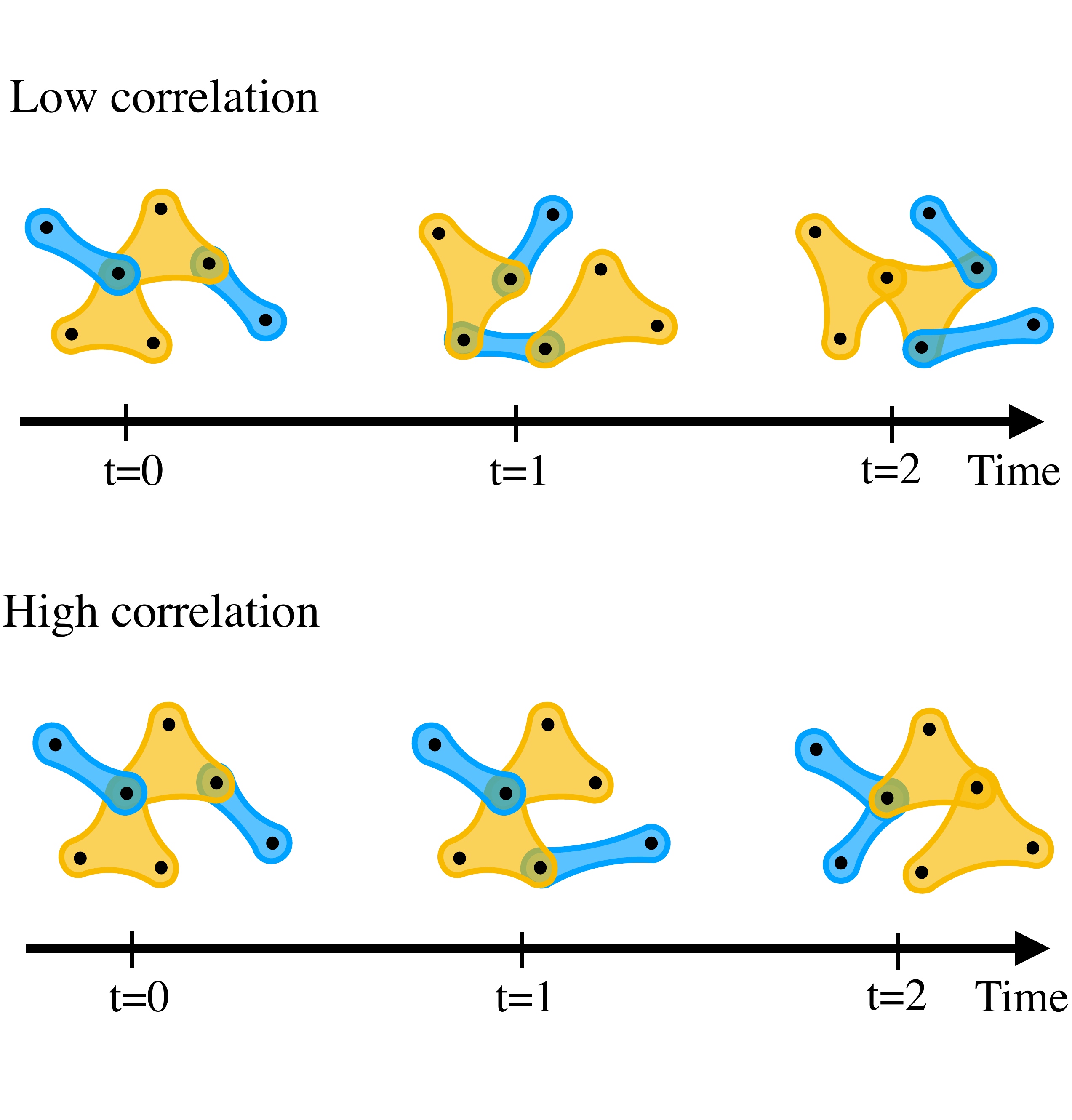}}
  \end{minipage}}
\usebox{\measurebox}\qquad
\begin{minipage}[b][\ht\measurebox][s]{.45\textwidth}
\centering
\subfloat
  []
  {\label{fig:corr_b}\includegraphics[width=1\textwidth]{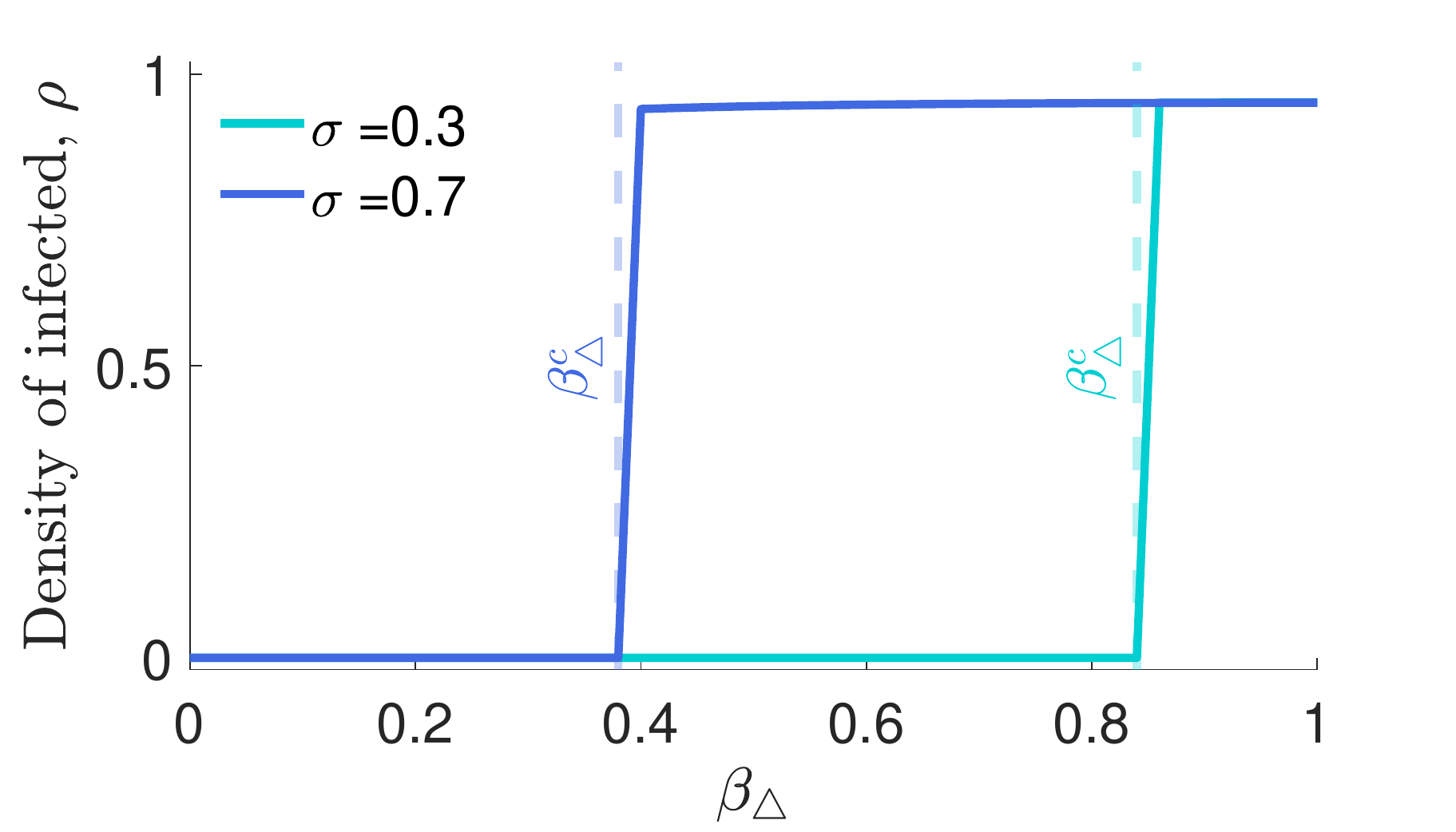}}
\vfill
\vspace{-0.3cm}
\subfloat
  []
  {\label{fig:corr_c}\includegraphics[width=1\textwidth]{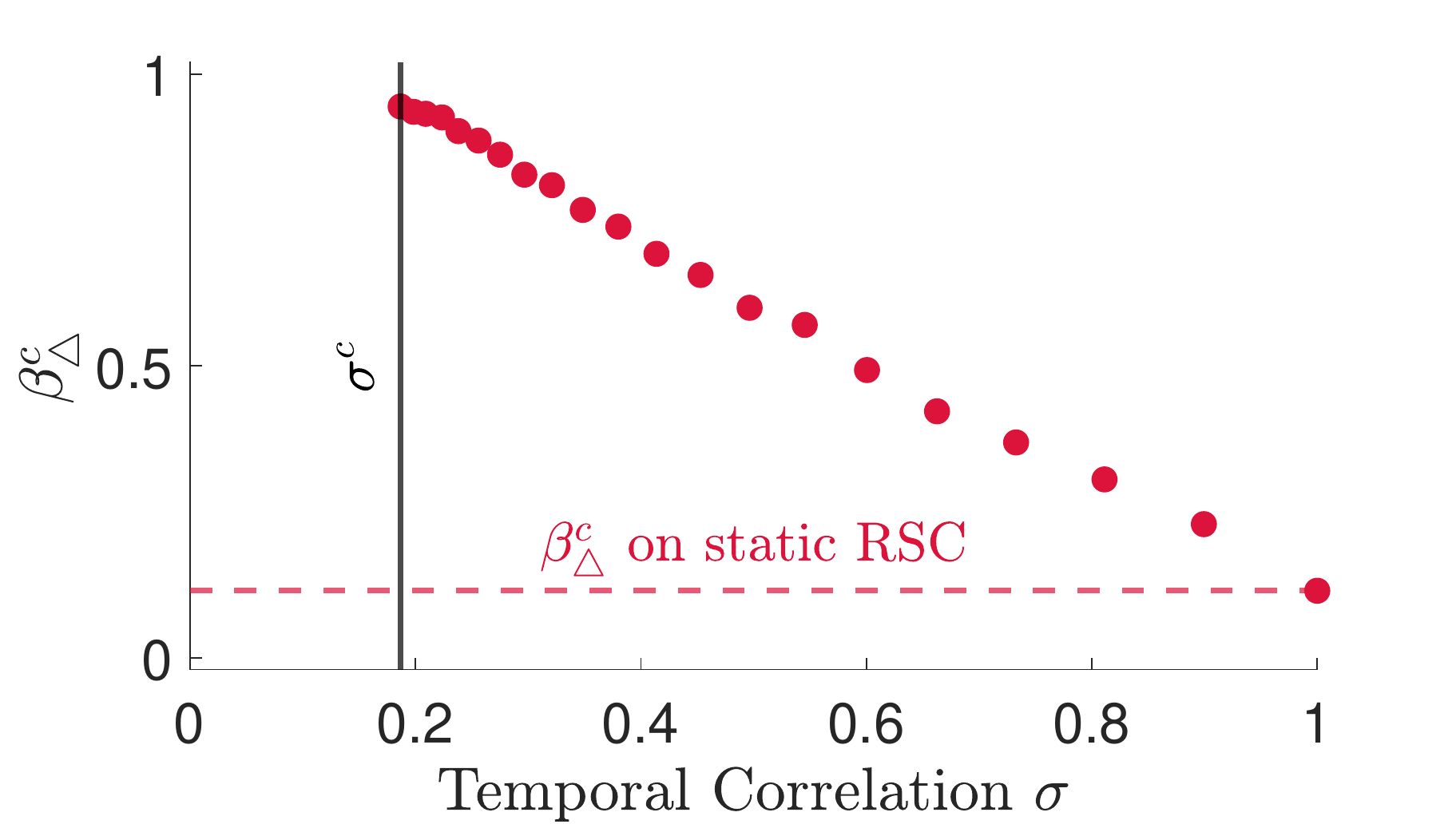}}
\end{minipage}
\vspace{0.8cm}
\caption{\textbf{Effect of temporal correlations in higher-order networks.} (a):  A schematic of temporal simplicial complexes with low and  high temporal correlations. (b): Size of the infected population at the steady state as a function of $\beta_\triangle$ for two different temporal correlations, $\sigma=0.7$ and $\sigma=0.3$. The critical value of the simplicial infection rate $\beta^c_\triangle$ to enter the endemic state is lower for higher temporal correlation. Both curves display an abrupt transition as a function of $\beta_\triangle$. (c): Critical simplicial infection rate computed as a function of the temporal correlation $\sigma$. The critical value $\beta^c_\triangle$ is higher for decreasing values of $\sigma$, and the epidemic threshold disappears below a critical value of temporal correlation $\sigma^c$ (grey line). This indicates that simplicial effects are stronger in highly correlated higher-order networks. For panels (b) and (c) we set $\beta_|=0.85\frac{\mu}{\langle{k_|}\rangle}$, and each point in (c) was obtained by averaging over 100 RSCs with $\langle{k_|}\rangle=12$ and $\langle{k_\triangle}\rangle=5$.}
\label{fig:corr}
\end{figure*}
In Fig.~\ref{fig:heat}, we fixed the size of the initial seed of infectious nodes at $\rho(0)=\frac{1}{N}$. To better characterize the two basins of attractions in the bistable regime and the associated critical mass effects, in Fig.~\ref{fig:finite} we vary the initial seed size and numerically investigate the onset of the epidemic. In particular, in Fig.~\ref{fig:finite_a} we first show the analytical solution for the stationary $\rho$ in the mean-field approximation derived in Ref.~\citep{iacopini2019simplicial} as function of $\lambda_|$ for different values of $\lambda_\triangle$. 
The dashed curves represent the unstable solutions that separate the basin of the infection-free state ($\rho=0$) from the endemic state ($\rho>0$). As $\lambda_\triangle$ is increased, we see that the basin of the infection-free state shrinks so that the endemic phase can be reached for progressively smaller values of initial infection size $\rho(0)$. 
Indeed, consistent with this, our numerical investigations on static simplicial complexes (Fig.~\ref{fig:finite_b}) reveal that while a small initial infection of size $p_0=\frac{0.1}{N}$ does not lead to early onset of endemic phase no matter the value of $\lambda_\triangle$,  increasing the initial seed size to $\frac{0.5}{N}$ or $\frac{1}{N}$ leads to early onsets on the endemic phase in our system with $N=500$ nodes. As expected, the onset occurs even earlier for higher values of $\lambda_\triangle$.
By contrast, in temporal simplicial complexes, as shown in Fig.~\ref{fig:finite_b}, the onset of the endemic phase in temporal simplicial complexes is largely independent of $\lambda_\triangle$, consistently with what was observed in Fig.~\ref{fig:heat_b}. This suggests that the basin of the infection-free state shrinks fast in static simplicial complexes as $\lambda_\triangle$ increases. As a consequence, the relevance of simplicial effects is strongly mitigated when we consider temporality, a realistic feature of many real-world social systems.

\subsection{Contagion on temporally correlated higher-order networks}
In the previous section we saw that introducing time-evolving structures can significantly impact contagion on higher-order networks, by altering the basin of the infection-free state in finite-size simplicial complexes. However, the way in which network structures evolve can be different. For instance, a social system may change more or less quickly, giving rise to different temporal correlations among networks at consecutive times. We thus consider as a measure of temporal correlation:
\begin{equation}
    \sigma= \frac{1}{2T}\sum_{t=1}^{T} 
\frac{n(|_t \cap |_{t+1})}{n(|_t \cup |_{t+1})} +\frac{n(\triangle_t \cap \triangle_{t+1})}{n(\triangle_t \cup \triangle_{t+1})}
\end{equation}
where $\triangle_t$ is the set of 2-simplices at time $t$ and $|_t$ is the set of 1-simplices which are not part of any 2-simplex at time $t$, $n(\triangle_t \cap \triangle_{t+1})$ is the number of 2-simplices that persist from time $t$ to the next time step $t+1$ and $n(\triangle_t \cup \triangle_{t+1})$  is the total number of 2-simplices present at time $t$ or $t+1$. Analogously, $n(|_t \cap |_{t+1})$ and $n(|_t \cup |_{t+1})$ are defined for 1-simplices. 

\begin{figure*}[!htb]
    \centering
    \begin{minipage}{.5\textwidth}
        \subfloat []
        {\label{fig:hetero_a}\includegraphics[scale=0.5]{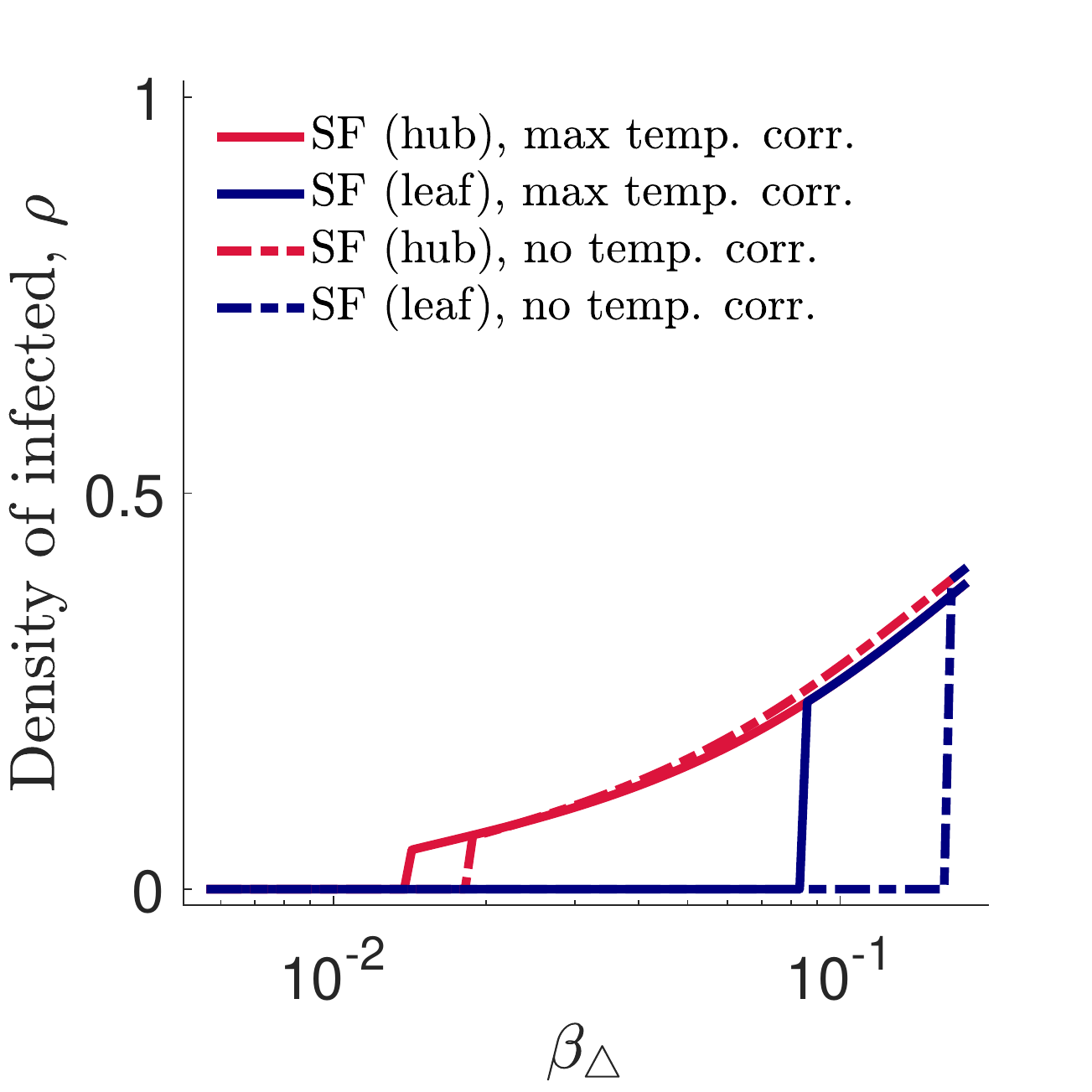}}
    \end{minipage}%
    \begin{minipage}{0.5\textwidth}
        \subfloat []
        {\label{fig:hetero_b}\includegraphics[scale=0.5]{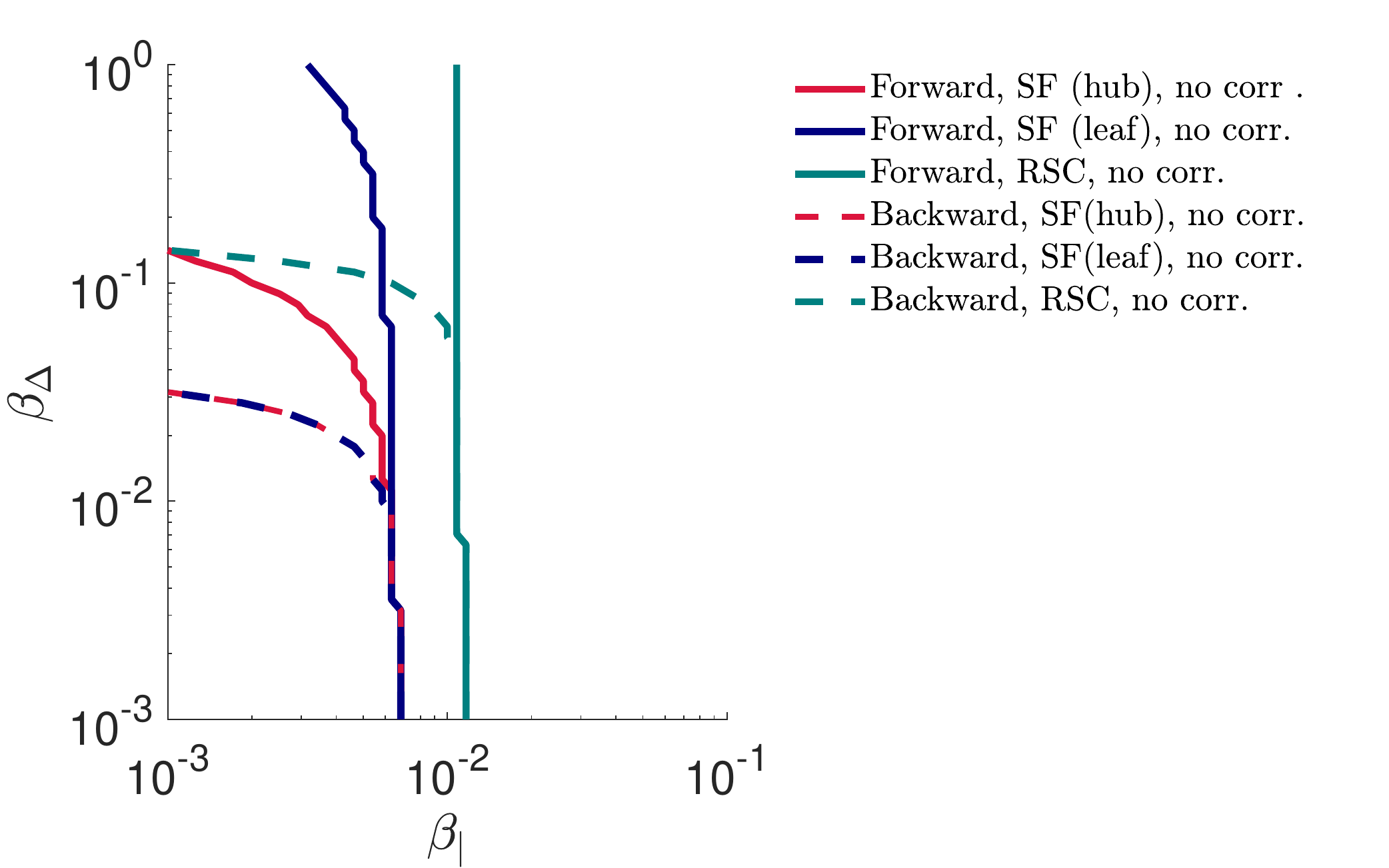}}
    \end{minipage}
   \caption{\textbf{Effect of heterogeneity in higher-order networks.} (a): Fraction of the infected population on heterogeneous scale-free (SF) simplicial complexes (power-law exponents $\gamma_|=2.2$ and $\gamma_\triangle=2.5$) as a function of $\beta_\triangle$ for maximum (dashed lines) and minimum (solid lines) temporal correlation. We consider two different scenarios for initial infection: hub (red) and leaf (blue). High temporal correlation reduces the epidemic thresholds. (b): Epidemic thresholds as a function of $\beta^c_|$ and $\beta^c_\triangle$ for heterogeneous and homogeneous simplicial complexes with the same number of interactions for the forward (solid curves) and backward (dashed curves) transition. On temporal SF complexes with no correlation, forward transition to the endemic state is possible for all considered values of $\beta_|$ upon increasing $\beta_\triangle$ both in the hub (red) and leaf (blue) seeding scenarios, in contrast with RSCs (green) where the lack of temporal correlation prevents the onset of endemic phase entirely below a critical $\beta_|$. For both SF and RSCs, the backward transition to infection-free state occurs upon decreasing $\beta_\triangle$, however a lower value of $\beta_\triangle$ is required for SF complex as compared to RSCs. For panel (a), we set  $\beta_|=0.25\frac{\mu}{\langle{k_|}\rangle}$, for both (a) and (b), we set $\mu=0.2$, $\langle{k_|}\rangle=10$, $\langle{k_\triangle}\rangle=4$.}
    \label{fig:hetero}
\end{figure*}

In order to investigate how the evolution of the network affects the spread of contagion, we introduce a model to systematically tune temporal correlations in simplicial complexes, where at each time the network is described by a RSC.
In details, we recursively generate a new simplicial complex at time $t+1$ by randomly rewiring with probability $f\in[0,1]$ the 1-simplices and 2-simplices present at time $t$. In this way, we are able to generate a temporal sequence of RSCs. Using such a model for sparse graphs, we can tune the temporal correlation $\sigma$ in an effective range between 0, describing the absence of correlation, and 1, where network structure does not change over time. Two schematics of temporal simplicial complexes with low and high correlation are shown in Fig.~\ref{fig:corr_a}.

In the following analysis, we focus on the forward transition to endemic state only, as the backward transition is unaffected by temporality as observed in Fig.~\ref{fig:heat} (dashed curves). We first infect a single node and simulate the epidemic process on top of two distinct sequences of temporal RSCs, one with correlation $\sigma=0.3$ and the other with correlation $\sigma=0.7$, and compute the fraction of infected nodes in the asymptotic state as a function of $\beta_{\triangle}$. As shown in Fig.~\ref{fig:corr_b}, in both cases the endemic phase is separated by an abrupt transition from the healthy region. The critical simplicial infection rate for the transition to occur is higher in the first case. 

We systematically investigate such phenomenon in Fig.~\ref{fig:corr_c}, where we compute the critical simplicial epidemic threshold as a function of $\sigma$. We observe that  $\beta_\triangle^c$ decreases monotonically with the temporal correlation $\sigma$ and it takes its minimum value for maximally correlated RSCs, corresponding to a static simplicial complex. Consistently with what was observed in Fig.~\ref{fig:heat} and Fig.~\ref{fig:finite}, this suggests not only that simplicial effects are weaker in temporal against a static setups, but that this is also the case the more diverse the temporal evolution of the system is.

We also note that the absence of a threshold $\beta_\triangle^c$ for values of temporal correlation below a critical $\sigma$, marked by a dashed vertical line, is due to the existence of a threshold of temporal correlation below which the transition to an endemic state is not possible, no matter the value of $\beta_\triangle$.

\subsection{Contagion on degree-heterogeneous temporal higher-order networks}

In the previous section we investigated the effects of temporality in homogeneous simplicial complexes. We now turn our attention to the role of degree heterogeneity in temporal higher-order networks~\cite{arruda2020social, landry2020effect, kovalenko2020growing, arruda2021phase}.

We generate scale-free (SF) simplicial complexes following a growth model introduced in \citep{kovalenko2020growing}, where both 1-simplices and 2-simplices follow a scale-free distribution, and where the sequences of $k_|$ and $k_\triangle$ are maximally correlated. Next, we obtain a temporal sequence of SF simplicial complexes via recursively performing degree preserved rewiring at each time step such that the degree distribution of the simplices does not change. Desired values of temporal correlation can be achieved by suitably choosing the rewiring probability. 

We simulate the epidemic process on top of two distinct sequences of SF simplicial complexes  corresponding to the two extreme values of temporal correlation $\sigma_{\rm{max}}=1$ and $\sigma_{\rm{min}}\approx 0$. For both configurations, we investigate two different scenarios of seeding infection, namely on the hub or on one of the leaves, and compute the fraction of the infected population in the long-time limit as a function of $\beta_\triangle$. As shown in Fig.~\ref{fig:hetero_a}, for both hub and leaf cases, the critical value of $\beta^c_\triangle$ to enter the endemic state is lower for higher values of temporal correlation, in agreement with what we found for homogeneous structures. Again, we only show the forward transition to the endemic state as the backward transition is not affected by temporality. As expected, seeding the infection on the hub enhances the epidemics. In particular, in the considered case, $\beta^c_\triangle$ decreases by an order of magnitude when the infection is started on the best connected node of the network.

To properly quantify the effect of heterogeneity, we systematically compare the onset of the endemic state in the heterogeneous simplicial complex as a function of both $\beta_|$ and  $\beta_\triangle$
against a homogeneous simplicial complex with the same number of 1- and 2-simplices. As shown in Fig.~\ref{fig:hetero_b}, in uncorrelated temporal SF complex, for the forward transition, it is possible to reach the endemic state for all $\beta_|$ below a critical value upon increasing $\beta_\triangle$. This is in contrast with RSCs where, below a critical $\beta_|$, the lack of temporal correlation prevents the onset of the endemic phase entirely, as already observed in Fig.~\ref{fig:heat_b}.
In such uncorrelated temporal case, for both SF and RSCs, the backward transition to infection-free state occurs upon decreasing $\beta_\triangle$, however a lower value of $\beta_\triangle$ is required for SF complex as compared to RSCs. Homogeneous structures are the safer against contagion: when structural heterogeneity is present, starting the epidemic from a peripheral node will have a milder effect than if contagion begins from the hub, but the system is more prone to reach the endemic state compared to a homogeneous network with the same number of interactions.

\section{Discussion}
In this work we have investigated the effect of temporality in spreading dynamics on higher-order networks. We focused primarily on the forward transition to the endemic state and showed that contagion processes behave remarkably differently on temporal and static finite-size homogeneous simplicial complexes. While in static networks the onset of the endemic state depends strongly on both $\beta_|$ and $\beta_\triangle$, in random temporal networks, where no correlations are present among time-consecutive interaction structures, the effect of the higher-order contagion parameter is much weaker. This is linked to changes in the basins of attractions of the epidemic-free state, which shrinks fast for static structures when increasing the infectivity of the 2-simplices.
As a consequence, temporality can have a direct impact on critical mass effects -- already present in the static case \cite{iacopini2019simplicial}-- by reshaping the basins of attractions of the system. In this scenario, a seed of infectious nodes of a fixed size can lead the system to both the endemic and epidemic-free states according to the temporal properties of its interactions.
More in details, we investigated the effect of the initial infection size on the onset of the endemic state, finding that while for very small values of initial infection the onset of the epidemic is not impacted by simplicial infectivity in both static and temporal simplicial complexes, a reasonable initial infection of size $\frac{1}{N}$ leads to striking differences between the two cases. Intermediate scenarios in the forward transition can be achieved on simplicial complexes with intermediate levels of temporal correlations. In contrast to the forward transition, we observed that the backward transition to infection-free state was unaffected by presence or absence of temporal correlations. 

We also investigated the effect of degree heterogeneity on higher-order contagion. We confirmed that even in scale-free simplicial complexes, the absence of temporal correlations increases the infectivity required to achieve the endemic phase. However, in contrast to homogeneous simplicial complexes, in heterogeneous structures the lack of temporal correlations does not completely hinder the effect of simplicial infectivity, and the endemic state can still be reached with a high enough value of $\beta_{\triangle}$. The parameter space associated to the endemic phase increases when the infection is seeded on a well-connected hub of the simplicial complex. However, even when the infection starts from a poorly connected node, the onset of the epidemics is always easier to achieve compared to an homogeneous simplicial complex with the same number of interactions.

Ref.~\cite{iacopini2019simplicial} first pointed out that higher-order interactions might lead to new emergent phenomena in spreading processes, for instance inducing new explosive contagion transitions which can not be achieved on traditional graphs where interactions are limited to dyadic ties. However, here we have shown that the early onset of such explosive transitions can be delayed in absence of temporal correlations, in some cases significantly reducing the parameter space associated to the emergence of the endemic state. As most higher-order social networks naturally evolve, with both pairwise and group interactions changing over time~\cite{cencetti2021temporal}, our results also suggest potential strategies to control contagion, by suitably tuning the temporal network structure.

Our work corroborates some ideas recently presented in \citep{st-onge2021bursty} on the importance of considering heterogeneity in disease modeling. There, the authors focus on
how bursty exposure to social environments (where the duration of higher-order interactions follows an exponential distribution) may affect contagion, showing through a mean-field analysis that the invasion threshold decreases with higher values of burstiness. In Ref.\citep{st-onge2021bursty} simplicial infectivity and burstiness are entangled together, and as a result simplicial infectivity is never independently or explicitly explored. Our framework of simplicial contagion, instead, allows us to explicitly disentangle temporality and simplicial infectivity. We use a different MMCA approach which operates at the level of single nodes, concluding that temporality may dominate higher-order effects in systems where both time-varying and group interactions are present.

In the future, our temporal framework could be applied to investigate other dynamical processes recently extended beyond pairwise interactions, including opinion~\cite{neuhauser2020multibody, hickok2021bounded}, convention~\cite{iacopini2021vanishing}, and evolutionary dynamics~\cite{alvarez2021evolutionary}. Taken together, our work suggests the importance to consider temporality, a feature of many real-world systems, when investigating dynamical processes on higher-order networks.

\section*{Acknowledgements}
\noindent A.K. acknowledges the Prime Minister's Research Fellowship  of the government of India for financial support. I.I. acknowledges support from the Agence Nationale de la Recherche (ANR) project DATAREDUX (ANR-19-CE46-0008). F.B. acknowledges partial support from the ERC Synergy Grant 810115 (DYNASNET).

\bibliographystyle{unsrtnat}

\end{document}